\documentclass[aip,graphicx]{revtex4-1}

\usepackage{amsmath, amsfonts, amssymb, mathrsfs, mathtools}
\usepackage{color, graphicx, subfigure, epstopdf}

\begin{document}

\title{String Theory has no Isotropic Solution for the Modified Einstein's Equations without the Dilaton}

\author{Christopher Frye}
\email[]{christopher.frye@knights.ucf.edu\\[-5pt]Address after August 1, 2013:\\[-5pt]Department of Physics, Harvard University,
                17 Oxford Street, Cambridge, MA 02138}
\affiliation{\mbox{Department of Physics, University of Central Florida, Orlando, FL 32816}}

\author{Costas Efthimiou}
\email[]{costas@physics.ucf.edu}
\affiliation{\mbox{Department of Physics, University of Central Florida, Orlando, FL 32816}}

\date{\today}

\begin{abstract}
We investigate the modification to Einstein's vacuum field equations which is imposed by string theory when the dilaton field is ignored. Including the cosmological constant in all calculations, we prove that such a theory of gravity admits no static isotropic solution. We then show that any isotropic solution of the equations in question must necessarily be static, therefore proving that no isotropic solution exists for this stringy modification to gravity.
\end{abstract}

\maketitle

\section{Introduction}

    In general relativity, one of the most powerful results is the existence of a static isotropic solution for Einstein's field equations
    --- the well known Schwarzschild metric --- which remains valid even when the static requirement is dropped. This extension is widely known as Birkhoff's theorem after Birkhoff presented it in his book\cite{Birkhoff}, although it was first discovered by
    Jebsen\cite{DF,Jebsen}. The isotropic solution to Einstein's equations is so useful that for any proposed modification of general relativity, we would like to know its analog.
    .

    String theory proposes one modification to general relativity.
    In the well known text of Green, Schwarz, and Witten\cite{GSW}, the modified Einstein's equations are given to be
    \begin{equation}
    \label{eq:modEinNoLambda}
        R_{\mu\nu} + \lambda\, R_{\mu\kappa\rho\tau}\,R_\nu^{~\kappa\rho\tau} = \mathcal{O}(\lambda^2)\,,
    \end{equation}
    where
    $\lambda \sim 10^{-70}$ m$^2$ is proportional to the inverse string tension\cite{Zwiebach} $\alpha'$.
    Though
    leaving out the dilaton field (which is not allowed\cite{Gasperini}), these equations are nonetheless interesting to study. We will examine a generalization of \eqref{eq:modEinNoLambda} which includes the cosmological constant:
    \begin{equation}
    \label{eq:modEin}
        R_{\mu\nu} - g_{\mu\nu} \, \Lambda + \lambda\, R_{\mu\kappa\rho\tau}\,R_\nu^{~\kappa\rho\tau} =
        \mathcal{O}(\lambda^2)\, .
    \end{equation}
    Obviously, any conclusions we make regarding \eqref{eq:modEin} apply also to \eqref{eq:modEinNoLambda} upon setting
    $\Lambda = 0$\,.

    In this paper, we prove that \eqref{eq:modEin} has no isotropic solution. We obtain this result by first attempting (and showing that it is impossible) to find a static isotropic solution to these equations; following the standard method\cite{Weinberg} used to find the Schwarzschild solution in general relativity works well for this step. Then we prove that any solution of \eqref{eq:modEin} which is isotropic must necessarily be static, thus completing our proof.

\section{No Static Isotropic Solution}

    To prove \eqref{eq:modEin} admits no \emph{static} isotropic solution, we plug the most general such metric\cite{Weinberg}
    \begin{equation}
    \label{eq:gen}
        ds^2 = -B(r)\,dt^2 + A(r)\,dr^2 + r^2\,d\theta^2+r^2\,\sin^2\theta\,d\phi^2
    \end{equation}
    into the left side of \eqref{eq:modEin}. Only the diagonal elements of the tensor on the left side do not vanish identically;
    they are\,:
    {\footnotesize
        \begin{align}
            \label{eq:tt}
            & R_{tt} ~+~ \Lambda\,B(r) ~+~ \lambda \, \left[ {-B'(r)^2 \over r^2\, A(r)^2\,B(r)} - { \Big\{ A(r)\,B'(r)^2 \,+\, B(r) \, \Big[ \, A'(r)\,B'(r) - 2\,A(r)\,B''(r)\, \Big] \Big\}^2 \over 8\,A(r)^4\,B(r)^3}~\right] \,,\\[15pt]
            \label{eq:rr}
            & R_{rr} ~-~ \Lambda\,A(r) ~+~ \lambda \, \left[ ~{A'(r)^2 \over r^2\,A(r)^3} + {\Big\{ A(r)\,B'(r)^2 \,+\, B(r) \, \Big[ \, A'(r)\,B'(r) - 2\,A(r)\,B''(r)\, \Big] \Big\}^2 \over 8\,A(r)^3\,B(r)^4}~ \right] \,, \\[15pt]
            \label{eq:thth}
            & R_{\theta\theta} ~-~ \Lambda\,r^2 ~ + ~{\lambda \over 2\,A(r)^4} \, \left[ ~{4 \, \big[ A(r) - 1\big]^2 \, A(r)^2 \over r^2} + A'(r)^2 + {A(r)^2\,B'(r)^2 \over B(r)^2} ~ \right] \,, \\[15pt]
            & R_{\phi\phi} ~-~ \Lambda\,r^2 \sin^2\theta ~ + ~{\lambda \sin^2 \theta \over 2\,A(r)^4} \, \left[ ~{4 \, \big[ A(r) - 1\big]^2 \, A(r)^2 \over r^2} + A'(r)^2 + {A(r)^2\,B'(r)^2 \over B(r)^2} ~ \right] \,;
        \end{align}
    }we call these expressions $(tt)$, $(rr)$, $(\theta\theta)$, and $(\phi\phi)$ respectively.  The nonvanishing components of
    the Ricci tensor $R_{\mu\nu}$ appearing here are given by
    \begin{align*}
        R_{tt} & ~=~ {B''(r) \over 2\,A(r)} - {A'(r)\,B'(r) \over 4\,A(r)^2} - {B'(r)^2 \over 4\,A(r)\,B(r)} + {B'(r) \over r\,A(r)}\,, \\[10pt]
        R_{rr} & ~=~ -{B''(r) \over 2\,B(r)} + {A'(r)\,B'(r) \over 4\,A(r)\,B(r)} + {B'(r)^2 \over 4\,B(r)^2} + {A'(r) \over r\,A(r)}\,,\\[10pt]
        R_{\theta\theta} & ~=~ 1 - {1 \over A(r)} + {r\,A'(r) \over 2\,A(r)^2} - {r\,B'(r) \over 2\,A(r)\,B(r)}\,,\\[10pt]
        R_{\phi\phi} & ~=~ R_{\theta\theta} \, \sin^2\theta\,.
    \end{align*}
    Right away, we see that we can disregard the $(\phi\phi)$ component of \eqref{eq:modEin} as redundant, leaving three equations for us to solve. Through \eqref{eq:modEin}, \eqref{eq:tt}, and \eqref{eq:rr} we also see that
    \begin{equation}
    \label{eq:addtwo}
        \mathcal{O}(\lambda^2) = {(rr) \over A(r)} + {(tt) \over B(r)} = {A'(r)\,B(r)+B'(r) A(r) \over r \, A(r)^2\,B(r)}
         + \lambda ~ {A'(r)^2 \,B(r)^2 - B'(r)^2\,A(r)^2 \over r^2\,A(r)^4\,B(r)^2}\,.
    \end{equation}
    At this point, we move from the general metric \eqref{eq:gen} to a more specific form. The de Sitter--Schwarzschild metric\cite{Rindler}
    \begin{equation}
    \label{eq:Schw}
        ds^2 ~=~ -\left(1-{2\,M\,G\over r} - {\Lambda\,r^2 \over 3}\right)\,dt^2 \:
                    +\: \left(1-{2\,M\,G\over r} - {\Lambda\,r^2 \over 3}\right)^{-1}\,dr^2 \:+\: \cdots
    \end{equation}
    solves \eqref{eq:modEin} to zeroth order in $\lambda$\,; we seek a static, isotropic solution to \eqref{eq:modEin} correct to first order in $\lambda$\,. This must thus have the form
    \begin{equation}
    \label{eq:ansatz}
        ds^2 \,=\, -\left[1 - {2MG\over r} - {\Lambda\,r^2 \over 3} + \lambda\,b(r) \right]\,dt^2
                    + \left[1 - {2MG\over r} - {\Lambda \, r^2 \over 3} + \lambda\,a(r)\right]^{-1} \,dr^2 + \cdots
    \end{equation}
    which amounts to setting
    \begin{equation}
    \label{eq:AB}
        B(r) = 1 - {2MG\over r} - {\Lambda\,r^2 \over 3} + \lambda\,b(r) \quad \text{and} \quad A(r)
                = \left[1 - {2MG\over r} - {\Lambda \, r^2 \over 3} + \lambda\,a(r)\right]^{-1}
    \end{equation}
    in \eqref{eq:gen}\,. Plugging these into \eqref{eq:addtwo}, we find that the second term on the right can be dropped since
    \[
        \lambda~ {A'(r)^2 \,B(r)^2 - B'(r)^2\,A(r)^2 \over r^2\,A(r)^4\,B(r)^2} ~=~ \mathcal{O}(\lambda^2)\,.
    \]
    Thus to enforce \eqref{eq:addtwo} we must set
    \begin{equation}
    \label{eq:const}
        A'(r)\,B(r) + A(r)\,B'(r) = \mathcal{O}(\lambda^2)\,, \quad \text{whence}
        \quad A(r)\,B(r) = 1 + \lambda\,k   + \mathcal{O}(\lambda^2) \,.
    \end{equation}
    We have written the constant of integration as $\,1 + \lambda k\,$, i.e.~as unity to zeroth order in $\lambda$\,; this comes by direct calculation using \eqref{eq:AB} or by requiring spacetime to be asymptotically Minkowskian. Now, \eqref{eq:AB} and \eqref{eq:const} imply that
    \begin{equation}
    \label{eq:ab}
        b(r) = a(r) + k\,\left(1 - {2\,G\,M \over r} - {\Lambda\,r^2 \over 3} \right)\,.
    \end{equation}
    With this relationship between $b(r)$ and $a(r)$ enforced, the $(tt)$ component of \eqref{eq:modEin} is simply a consequence of the $(rr)$ component. We are thus left with just the $(rr)$ and $(\theta\theta)$ components of \eqref{eq:modEin} as independent equations. Beginning with the $(\theta\theta)$ equation, we use \eqref{eq:AB} and \eqref{eq:ab} in \eqref{eq:thth} which yields
    \[
        r \, a'(r) + a(r) = {12\,G^2\,M^2 \over r^4} + {2\,\Lambda^2\,r^2 \over 3}
    \]
    with solution
    \begin{equation}
    \label{eq:a}
        a(r) = {c \over r} - {4\,G^2\,M^2 \over r^4} + {2\,\Lambda^2\,r^2 \over 9}\,,
    \end{equation}
    keeping the constant of integration $c$ arbitrary for now. Through \eqref{eq:ab} this determines
    \begin{equation}
    \label{eq:b}
        b(r) = {c \over r} - {4\,G^2\,M^2 \over r^4} + {2\,\Lambda^2\,r^2 \over 9} + k\,\left(1 - {2\,G\,M\over r} - {\Lambda\,r^2 \over 3}\right)
    \end{equation}
    which should conclude the calculation. However, we must check to see whether this solution is consistent with the equation $(rr) = 0$\,. Plugging \eqref{eq:AB} into \eqref{eq:rr} using \eqref{eq:a} and \eqref{eq:b} gives
    \[
        (rr) = {36\,\lambda \over r^4} \, \left({G\,M\over r}\right)^2 \, \left(1 - {2\,G\,M \over r} - {\Lambda\,r^2 \over 3}\right)^{-1} \neq 0\,.
    \]
    It thus appears we have an inconsistent system of differential equations. Notice the inconsistency disappears in the classical limit as $\lambda \to 0$\,.

\section{Analog to Birkhoff's Theorem}

    We can push this result a step further by proving that, in fact, no isotropic solution of \eqref{eq:modEin} exists, dropping the static condition. We do this by showing that any isotropic solution of \eqref{eq:modEin} must necessarily be static. That is, we prove an analog of Birkhoff's theorem for this modified theory of gravity.

    The most general isotropic metric can be written
    \[
        ds^2 = -B(r,t)\,dt^2 + A(r,t)\,dr^2 + r^2\,d\theta^2 + r^2\,\sin^2\theta\,d\phi^2\,.
    \]
    Since we know that Birkhoff's theorem\cite{Weinberg} holds true to zeroth order in $\lambda$ (presence of $\Lambda$ does not disrupt the proof), any isotropic solution to \eqref{eq:modEin} can be written as
    \[
        ds^2 = -\left[1-{2\,M\,G\over r} - {\Lambda\,r^2 \over 3}+ {\lambda}\,b(r,t) \right]\,dt^2 + \left[1-{2\,M\,G\over r} - {\Lambda\,r^2 \over 3} + {\lambda}\,a(r,t)\right]^{-1} \,dr^2 +\cdots\,;
    \]
    this defines $A(r,t)$ and $B(r,t)$ more specifically.
    Plugging this metric into the left side of \eqref{eq:modEin} gives some nonvanishing off-diagonal components, e.g.
    \[
        \mathcal{O}(\lambda^2) = (tr)
          = - {\lambda \, \dot a(r,t) \over r} \, \left( 1 - {2\,G\,M \over r} - {\Lambda\,r^2 \over 3} \right)^{-1}\,;
    \]
    here a dot denotes differentiation with respect to time and a prime differentiation with respect to the radial coordinate. This equation is very convenient, as it implies $a = a(r)$ and $A = A(r)$\,, thus simplifying remaining equations. Now expanding
    \begin{align*}
        \mathcal{O}(\lambda^2) &= {(tt) \over B(r,t)} + {(rr) \over A(r)} \\[10pt]
                   &= {\lambda \over r} \, \left(1 - {2\,G\,M \over r} - {\Lambda\,r^2 \over 3}\right)^{-1} \,
                   \left\{ ~ \Big[a'(r) - b'(r,t)\Big] \, \left( 1 - {2\,G\,M \over r} - {\Lambda\,r^2 \over 3} \right) \right. \\[10pt]
                   & \qquad \qquad \qquad \qquad \qquad \qquad \qquad ~~  \left. +~ \Big[ a(r) - b(r,t)\Big] \,
                   \left( {2\,G\,M \over r^2} - {2\,\Lambda\,r \over 3} \right) ~ \right\}
    \end{align*}
    we arrive at the equation
    \[
        A'(r)\,B(r,t) + A(r)\,B'(r,t) = \mathcal{O}(\lambda^2)\,;
    \]
    this implies $A(r)\,B(r,t) = f(t)$\,, for some function $f$ of $t$. But no matter the functional form of $f(t)$, through a change of coordinates in which
    \[
        dt' = {dt \Big / f(t)}\,,
    \]
    this function may be absorbed into $B(r,t)$\,. Let us assume we began the calculation in such a coordinate system (so we can avoid an overflow of new symbols)\,; we are then left with $A(r)\,B(r,t) = $ const. This can only hold when $B = B(r)$\,, leaving us with a static field.

\section{Conclusion}
\label{sec:conc}

    We have shown that  the dilaton-free stringy modification to Einstein's field equations admits no isotropic solution. Physically, this
    result is somewhat surprising. Indeed, Einstein's field equations do admit an isotropic solution, but upon perturbation by the small term $\lambda\, R_{\mu\kappa\rho\tau}\,R_\nu^{~\kappa\rho\tau}$, which does not pick out a preferred direction in spacetime, the solution disappears instead of being perturbed. However, mathematically this is not such a great surprise: we attempted to solve a set of three differential equations with two unknown functions. In classical gravity, one can plug the general metric \eqref{eq:gen} into $\,R_{\mu\nu} - \Lambda\,g_{\mu\nu} = 0\,$ and find the de Sitter--Schwarzschild solution. In that case, after determining that $A(r) = 1/B(r)$ (i.e.~our equation \eqref{eq:const} with $\lambda$ set to zero), one can determine that
    \[
        \big( R_{rr} - \Lambda\,g_{rr} \big) - {1 \over 2\,r\,B(r)}\,{d \big( R_{\theta\theta} - \Lambda\,g_{\theta\theta} \big) \over dr} = 0\,,
    \]
    reducing the number of independent equations to two. In our problem, if we plug
    \[
        A(r) = {1+\lambda\,k \over B(r)}
    \]
    from \eqref{eq:const} above into \eqref{eq:rr} and \eqref{eq:thth} we find that
    {\small
        \begin{align*}
            (rr) - {1 \over 2\,r\,B(r)}\,{d(\theta\theta) \over dr} ~=~ {\lambda \over 2\,r^4\,B(r)} ~ & \Bigg\{ 4 + 4\,B(r)^2 + 2\,r^2\,B'(r)^2 - 4\,B(r)\,\big[ r\,B'(r)+2 \big] \\ &- k\,r^4\,B''(r) + r^4\,B''(r)^2-2\,r\,B'(r)\,\big[ r^2\,B''(r) + k\,r^2 - 2\big] \Bigg\}\,,
        \end{align*}
    }which does not vanish but is of order $\lambda$\,. In contrast to the classical problem of determining an isotropic solution to Einstein's equations, our system in the perturbed problem is overdetermined.
    This problem is eliminated if the dilaton field is allowed in the equations.
    
    The first-order  correction to Einstein's vacuum field equations, including the necessary dilaton-graviton coupling, are\cite{BlackHoles}
    \begin{align*}
        R_{\mu\nu} + 2 \, \nabla_\mu \, \nabla_\nu \, \Phi
                        + \lambda \, R_{\mu\alpha\beta\gamma}\,R_\nu^{~\alpha\beta\gamma} ~=&~~
                         \mathcal{O}(\lambda^2)  \, , \\[5pt]
        \square \Phi - (\nabla \Phi)^2 + {1 \over 4} \, R
                        + {1 \over 8} \, \lambda \, R_{\alpha\beta\gamma\delta} \, R^{\alpha\beta\gamma\delta} ~=&~~
                        \mathcal{O}(\lambda^2) \,  .
    \end{align*}
    These equations have been studied extensively, and their static isotropic solution is known\cite{BlackHoles}:
    \begin{align*}
        \Phi ~&=~ - {2\,G\,m \over 3\,r^3} - {1 \over 2\,r^2} - {1 \over 2\,G\,m\,r} \,, \\[5pt]
        ds^2 ~&=~ - B(r) \, dt^2 + A(r) \, dr^2 + r^2\,d\theta^2 + r^2 \, \sin^2 \theta \, d\phi^2 \,,
    \end{align*}
    with
    \begin{align*}
        B(r) &= \left( 1-{2\,G\,m \over r} \right) \; \Bigg[ 1 - {2\lambda \over r^2} \, \left(
        {23\,r \over 24\,G\,m} + {11 \over 12} + {G\,m\over r}
        \right) \Bigg]\;, \\[15pt]
        A(r) &= \left(1-{2\,G\,m \over r} \right)^{-1} \; \Bigg[ 1 - {2\lambda \over r^2} \, \left(
        {r \over 24\,G\,m} + {7 \over 12} + {5\,G\,m \over 3\,r}
        \right) \Bigg]\,;
    \end{align*}
    obviously, we cannot set $\Phi = 0$ to get a sans-dilaton solution.
    In another paper\cite{FE} we use this isotropic solution to compute the corrections string theory predicts to the classical tests of
    general relativity.

\begin{acknowledgments}
CF expresses his deepest gratitude to his undergraduate advisor CE for providing guidance during this project. Both authors 
thank  Tristan H\"ubsch for reading the initial manuscript and providing  feedback.
\end{acknowledgments}

\end{document}